\documentclass[letter]{aa}  

\usepackage{graphicx}
\usepackage{txfonts}
\begin{document}

   \title{The role of supernovae inside AGN jets in UHECR acceleration}

\titlerunning{SNe within AGN jets and UHECRs}

   \author{V. Bosch-Ramon
          \inst{1}
          }

\institute{Departament de F\'{i}sica Qu\`antica i Astrof\'{i}sica, Institut de Ci\`encies del Cosmos (ICC), Universitat de Barcelona (IEEC-UB), Mart\'{i} i Franqu\`es 1, E08028 Barcelona, Spain. \\\\  \email{vbosch@fqa.ub.edu}
             }

 
  \abstract
   {Jets of active galactic nuclei are potential accelerators of ultra high-energy cosmic rays. Supernovae can occur inside these jets and contribute to cosmic ray acceleration, particularly of heavy nuclei, but that contribution has been hardly investigated so far.}
   {We carried out a first dedicated exploration of the role of supernovae inside extragalactic jets in the production of ultra high-energy cosmic rays.}
   {We characterized the energy budget of supernova-jet interactions, and the maximum possible energies of the particles accelerated in those events, likely dominated by heavy nuclei. This allowed us to assess whether these interactions can be potential acceleration sites of ultra high-energy cosmic rays, or at least of their seeds. For that, we estimated the cosmic ray luminosity for different galaxy types, and compared the injection rate of cosmic ray seeds into the jet with that due to galactic cosmic ray entrainment.}
   {Since the supernova is fueled for a long time by the luminosity of the jet, the energy of a supernova-jet interaction can be several orders of magnitude greater than that of an isolated supernova. Thus, despite the low rate of supernovae expected to occur in the jet, they could still provide more seeds for accelerating ultra high-energy particles than cosmic ray entrainment from the host galaxy. Moreover, these interactions can create sufficiently efficient accelerators to be a source of cosmic rays with energies $\gtrsim 10$~EeV.}
   {Supernova-jet interactions can contribute significantly to the production of ultra high-energy cosmic rays, either directly by accelerating these particles themselves or indirectly by providing pre-accelerated seeds.}

   \maketitle

\section{Introduction}\label{intro}

For a long time, the jets of active galactic nuclei (AGNe) have been considered possible sources of
ultra high-energy cosmic rays (UHECRs) (\citealt{hil84}; see also \citealt{rie22} and
\citealt{mat23} for recent reviews). In fact, extragalactic radio jets were already viewed as
likely sources of cosmic rays (CRs) much before the nature of these structures was well understood
\citep[e.g.,][]{bur62}, and in the last decades different jet regions and jet-medium interaction
sites have been proposed to be efficient accelerators of UHECRs
\citep[e.g.,][]{rac93,rom96,ost98,aha02,der07,mat19,rie22,zir23}. Observations hint at
starburst galaxies as somewhat more likely UHECR sources than AGN jets \citep[e.g.][]{dim23},
although there is debate on the former's capability to accelerate nuclei up to those energies
\citep[e.g.,][]{anc18,rom18}, and both source types may in fact be needed to explain observations in time-dependent
scenarios \citep[e.g.,][and references therein]{tay23}. 

An important requirement for jets as UHECR accelerators is that the magnetic power of the jet
must be sufficient to accelerate and confine ultra high-energy (UHE) charged particles.
This power requirement is greatly reduced if the highest energy particles are significantly heavier than H or He
\citep[e.g.,][]{lem09,rie22}. Experimental results seem to indicate that UHECR nuclei are on average
rather heavy (around CNO masses) at energies $\gtrsim 10$~EeV \citep[e.g.,][]{aab14,pie19}. This high
metallicity, and the observed spectrum, may require that those UHECRs are accelerated out of a pool of
rather heavy nuclei, and with a quite hard spectrum if source variability is not relevant
\citep[e.g.,][]{tay15,abd23}. If UHECR acceleration takes place within or around extragalactic jets,
large amounts of heavy ions can reach the acceleration sites through winds of evolved stars
\citep[e.g.,][]{wyk13,wyk15,wyk18} or supernova (SN) ejecta \citep[e.g.,][]{vie19,tor19} inside the
jet. Both of these channels are related, but the latter has not been discussed in detail in the
context of UHECR production. Diffusive entrainment of heavy high-energy galactic cosmic rays from the host galaxy (GCRs) could be another possibility \cite[e.g.,][]{cap15,kim18,mba19,seo22}. 
Large particle mean free
paths are often needed for efficient acceleration, either to cross the jet-medium shear layer
and/or to sample the internal jet velocity structure \citep[e.g.,][]{rie22}, so UHECR seeds may
have to be quite energetic to engage in processes capable of accelerating them
to UHE. For the aforementioned diffusely entrained GCRs, only those able to penetrate deep into the jet could also have the appropriate energies to participate in further acceleration
\citep{cap15}. On the other hand, for matter shed by SNe directly inside the jet, the associated interaction region could be an efficient accelerator itself (e.g., \citealt{vie19}; for evolved stars, see, e.g.,
\citealt{bar10,tor19b}), and nuclei accelerated there may already reach UHE, or serve as
pre-accelerated seeds for UHECR acceleration elsewhere in the jet (shear layers, shocks, etc.) and its
termination region and subsequent backflow shocks \citep{mat19,cer23}. 

In this work, to better understand the role of SNe occurring inside AGN jets in the production of
UHECRs, we carried out a first exploration of these events either as UHECR acceleration sites, or as
providers of pre-accelerated seeds. We did not focus on a specific AGN host type, as SNe of
different types are expected in different sorts of galaxies. The convention $A_x=(A/10^x\,{\rm
cgs})$ was adopted unless otherwise stated.

\section{Supernovae in extragalactic jets}\label{sne}

Evolved stars or compact stellar systems cross an extragalactic jet in $\sim r_{\rm
j}/v_\star\approx 10^6r_{\rm j,100pc}\,v_{\star,7}^{-1}$~yr, adopting a reference distance $\lesssim 1$~kpc from the galactic
plane (see below). These objects, unless very close to the base of the jet, should evolve unaffected by it because supersonic winds prevent the jet from reaching them
\citep[e.g.,][]{bar10,kha13,ara13,per17}. This is particularly true for the regions where most
core-collapse (CC) and type~Ia SNe take place. On those scales, a B0V-type star with wind mass rates
and velocities $\sim 10^{-8}$~M$_\odot$~yr$^{-1}$ and $\sim 10^8$~cm~s$^{-1}$ \citep{krt14}, very
conservative for stars close to becoming CC~SNe, would create a bow-shaped interaction region at $\sim
(3-30)\times 10^3\,L_{\rm j,44}^{-1/2}$~$R_*$ from the star, where $R_*\approx 10\,$R$_\odot$ is the
stellar radius, $L_j$ the jet power, and $\theta\sim 0.1$~rad the jet half-opening angle. Even for a
solar-type wind one gets an interaction distance of $\sim (0.2-2)\times 10^2\,L_{\rm
j,44}^{-1/2}$~R$_\odot$, which is a conservative estimate in type~Ia SN
single-degenerate progenitors as strong winds are expected from an evolved donor or white dwarf
accretion; whereas in the very compact double-degenerate progenitors, jet influence is unlikely
\citep[see][and references therein]{liu23}. Thus, both CC and Type~Ia SNe should occur inside the
jet. 

The possibility of SNe exploding inside extragalactic jets have been occasionally considered in the
literature \citep[e.g.,][]{bla79,fed96,bed99,vie19,tor19}, but its consequences have been very
rarely explored in detail. For instance, \cite{vie19} show that CC~SNe, exploding at 50~pc
from the jet base in an AGN hosted by a starburst galaxy, could produce detectable high-energy
emission from some sources in the local Universe.  It is worth noting however that CC~SNe
inside AGN jets are also expected in late-type hosts with moderate star formation on scales up to
hundreds of parsecs from the galactic plane, whereas Type~Ia SNe are expected in both
late-type and early-type galaxies, with a larger spatial spread of approximately a kiloparsec
\citep[e.g.,][]{geb09,hak17,cro21}. For early-type galaxies, \cite{tor19} propose that
type~Ia SNe occurring inside the jet could also lead to
detectable high-energy radiation. To our knowledge though, the role of jet-ejecta interactions in UHECR
production has not been explored so far, with the exception of a brief discussion
in \cite{bed99}. Given that CC and type-Ia SN yields are either dominated by oxygen or iron,
respectively \citep[see, e.g., table~2 in][]{day18}, they could play an important role in
explaining the medium-to-heavy composition of UHECRs at $\gtrsim 10$~EeV. We note that SN
remnants (SNRs) in the galaxy hosting an ANG jet, but located outside the jet, have already been considered to explain the UHECRs, which is a scenario complementary to ours \citep[e.g.,][if SNRs are behind the highest energy GCR]{cap15}.

In this work, we focus on SNe despite the fact that total mass shed by evolved stars in the jet should surpass
that released by SNe \citep[e.g.,][]{hub06}. As shown below, the ejecta mass and release time
are large and short enough, respectively, so the shocked ejecta can eventually cover the whole jet
cross section while producing a relatively coherent interaction structure. This would be hard to achieve
by a more gradual mass injection, although a proper assessment requires detailed investigation since
the rare and strong mass-loss events that evolved stars can suffer might have similar consequences. 

\section{Jet-ejecta interaction energetics}

For our exploration of the jet-ejecta interaction scenario, we took $L_{\rm
j}=10^{44}$~erg~s$^{-1}$ as the reference value for the jet power because significantly less
powerful jets should not be able to accelerate UHECRs to the highest energies \citep{lem09}. On the
other hand, significantly more powerful jets are very rare in the local Universe \citep[see, e.g.,
Fig.~4 in][]{min19}, and UHECRs cannot come from significantly beyond 100~Mpc
\citep[e.g.,][]{gre66,zat66}. Thus, $L_{\rm j}=10^{44}$~erg~s$^{-1}$ is a reasonable choice
\citep[it incidentally coincides with the FRI/FRII radio galaxy divide; e.g.,][and references
therein]{per19}.

We adopted a simplified SN model with an energy reference value of $E_{\rm SN}=10^{51}$~erg \citep[e.g.,][]{lea17,lea22}, 
mostly in kinetic form, and homologously expanding as a sphere of uniform density and radius $R_{\rm
ej}$. The mass of the SN~ejecta was fixed to $M_{\rm ej}=2$~M$_\odot$, in between typical type~Ia and
average CC SN values \citep{day18}. The SN~ejecta in reality is expected to present a strong density drop
beyond a radius with flow velocity $\sim\sqrt{E_{\rm SN}/M_{\rm ej}}$ \citep[slightly different depending on the SN
type; see, e.g.,][and references therein]{pet21}, but in the adopted model of unshocked ejecta the velocity at maximum radius is the following:  
\begin{equation} 
v_{\rm exp}\approx \sqrt{10E_{\rm
SN}/3M_{\rm ej}}\approx 9.1\times 10^8 E_{\rm SN,51}^{1/2}(M_{\rm SN}/2\,{\rm M}_\odot)^{-1/2}\,{\rm
cm~s}^{-1}. 
\end{equation}

The SN explosion was assumed to occur in a jet region with radius $r_{\rm j}=0.1$~kpc, which for a jet with $\theta\sim 0.1$~rad implies a distance $z_{\rm j}\sim 1$~kpc from the jet base. Nevertheless, the
relevant quantity here is $r_{\rm j}$ because it determines the jet ram pressure in the lab frame $P_{\rm j}$, and $z_{\rm j}\sim 1$~kpc is useful just as a reference. The jet is expected to be moderately relativistic there \citep[e.g.,][]{mul09,per14,red23} since the jet may have endured different sorts of external and internal dissipative processes already \citep[e.g.,][]{per19}, being still supersonic but relatively hot, mass-loaded, and slowed down. Given the uncertainty, for simplicity $P_{\rm j}$ was taken $\approx L_{\rm j}/\pi r_{\rm j}^2c$.

The impact of the jet becomes relevant for the dynamics of the unshocked expanding ejecta when their ram pressures become similar (in the lab frame) at the ejecta side facing the
jet. The jet stops the expansion on that side of the SN~ejecta when the latter radius is as follows:
\begin{equation}
R^0_{\rm ej}\approx (5E_{\rm SN}/2\pi P_{\rm j})^{1/3}\approx 13.4\,E_{\rm SN,51}^{1/3}L_{\rm 44}^{-1/3}\,{\rm pc}\,.
\end{equation}
At this stage, the strong shock driven into the SN~ejecta by the jet ram pressure moves with velocity $v_{\rm s}\approx v_{\rm exp}$ in the lab frame, derived from $\rho_{\rm ej}v_{\rm s}^2\approx P_{\rm j}$, where $\rho_{\rm ej}$ is the unshocked ejecta density\footnote{We neglect the velocity distribution of the unshocked ejecta, which should affect the shock speed only in a small fraction of ejecta volume.}. Later on, $v_{\rm s}$ grows because the pre-shock $\rho_{\rm ej}$ drops due to the unshocked ejecta expansion. The ejecta shock is equivalent to that in SNRs expanding in the interstellar medium, but here it is fueled by the jet impact. 

Given that the unshocked ejecta still expands with velocity $v_{\rm exp}$, at this stage the actual jet cross section covered by the ejecta grows with time as $\sigma_{\rm ej}\propto (R_{\rm ej}/R^0_{\rm ej})^2=(1+\hat x)^2$, with $\hat x=v_{\rm exp}t/R^0_{\rm ej}$. On the other hand, $\rho_{\rm ej}$ in the unshocked region homologously falls as $\propto 1/(1+\hat x)^3$, so $v_{\rm s}\propto (1+\hat x)^{3/2}$. Finally, one must account for the expansion of the ejecta in the direction away from the jet-driven shock, increasing its length in that direction as $\propto (2+\hat x)$. From all of this, the total jet-driven shock luminosity and total energetics are
$L_{\rm s}\sim \sigma_{\rm ej}\rho_{\rm ej}v_{\rm s}^3/2\propto (1+\hat x)^{7/2}$ and $E_{\rm s}\sim\int_0^{t_{\rm max}}(1/2)\sigma_{\rm ej}\rho_{\rm ej}v_{\rm s}^3 dt$,
respectively, where $t_{\rm max}$ can be derived from $\int_0^{t_{\rm max}}v_{\rm s}dt\sim 2R_0+v_{\rm exp}t_{\rm max}$. These equations lead to $\hat x\sim 1.5$, $v_{\rm s}\approx 4\,v_{\rm exp}$, and $E_{\rm s}\sim 17E_{\rm SN}$ at $t_{\rm max}\approx 1.5\,R^0_{\rm ej,10pc}v^{-1}_{\rm exp,9}$~kyr. Thus, the energy of the jet-driven shock crossing the ejecta is already significantly larger than $E_{\rm SN}$, but as subsequently shown the whole jet-ejecta interaction energy, part of which may go to UHECR acceleration, may be much higher.

The jet is also shocked when colliding with the ejecta. The energy dissipating in the jet shock while
the ejecta is being shocked is $\approx v_{\rm j}/v_{\rm s}\approx 10\,v_{\rm s,9.5}^{-1}$ times
higher than in the ejecta shock for $v_{\rm j}\sim c$. The jet composition is in principle much lighter than in the
SN~ejecta, dominated by jet protons, $e^\pm$, and H and He (mostly) entrained from the jet immediate
vicinity. Therefore, the nuclei accelerated by the jet shock might be too light to be behind a heavy UHECR component, although this may not be true because heavy GCR can reach the jet upstream
of the shock \citep[as in][]{cap15}, or ejecta matter can reach the jet shock due to mixing in later stages of the interaction (see below).

The whole jet-ejecta interaction is longer than $t_{\rm max}$, and strong jet energy dissipation can
feed CR production as long as the ejecta material stays in the path of the jet and their relative
velocity is high enough \citep{vie19}. Two possibilities can be realized: (i) The shocked ejecta
could be accelerated up to a Lorentz factor $\sim \Gamma_{\rm j}$, which would potentially involve as
much energy as $E_{\rm s}\sim \Gamma_{\rm j}M_{\rm ej}c^2\approx 10^{55}\Gamma_{\rm j,0.5}(M/2\,{\rm
M}_\odot)$~erg. Although a detailed characterization of the whole jet-ejecta interaction needs a
thorough numerical investigation, which is left for future work, 2D relativistic hydrodynamic simulations
\citep{vie19} already show that the SN~ejecta eventually covers the whole jet cross section and
quickly gets disrupted. Some jet-ejecta mixing can occur. Since the jet is surrounded by a much
denser medium, the latter effectively confines the disrupting ejecta, which is still pushed along the
jet channel. (ii) The effective acceleration of the SN ejecta may however stop before it
gets relativistic if the jet got disrupted as well. In that case, the medium, ejecta, and jet would get
temporarily mixed, and the fresh jet material would have to clear its way before recovering a more or
less steady configuration. The timescale in the lab frame of the whole process ($t_{\rm dyn}$) in
both (i) and (ii) is $\gtrsim t_{\rm max}+z_{\rm j}/v_{\rm j}$, as the ejecta has to expand,
accelerate, and/or be pushed aside \citep[see, e.g.,][for the acceleration phase]{bar12,kha13}.
Remarkably, even if both the ejecta and the jet became disrupted at the interaction region in (ii), the fresh
jet flow could still suffer a strong shock due to a sudden deceleration \citep{vie19}.
Taking $t_{\rm dyn}\sim 10^4$~yr and $L_{\rm j,44}\sim 1$ yields an estimate of the involved energy
in (ii) of $E_{\rm s}\sim t_{\rm dyn}L_{\rm j}\approx 3\times 10^{55}t_{\rm dyn,11.5}L_{\rm
j,44}$~erg. 

\section{CR maximum energy in jet-ejecta interactions}\label{acc}

Nuclei accelerated in the jet-ejecta interaction can reach an energy $E_{\rm max}$ and escape the region in a time $<3\times 10^4z_{\rm j,1kpc}\,v_{j,9.5}^{-1}$~yr before cooling or disintegrating via synchrotron, $\gamma$-meson production, $e^\pm$ creation, and photodisintegration \citep[e.g.,][]{hil84,mat23}: given the cross sections of $\sim 0.1-10$~mb \citep[synchrotron is negligible;][]{rac96,aha02b,kel08}, and adopting a local IR photon density of $n_{\rm IR}\sim 10^3\,L_{\rm IR,44}/z_{\rm j,kpc}^{-2}$~cm$^{-3}$ (also $\sim n_{\rm CMB}$), the cooling and disintegration times are $> 10^5$~yr. Assuming diffusive shock acceleration in the Bohm regime, taking reference values for the parameters typical for the ejecta shock, and a lab frame magnetic field $B\propto B_{\rm eq}$ (where $B_{\rm eq}^2/4\pi=P_{\rm j}$, with $B_{\rm eq}\sim 1\,L_{\rm j,44}r_{\rm j,100pc}^{-1}$~mG), one obtains the following:
\begin{equation}
E_{\rm max}\sim 4\,Z_1 R_{\rm ej,10pc}v_{\rm s,9.5}(B/B_{\rm eq})L_{\rm 44}^{0.5}r_{\rm j,100pc}^{-1}\,{\rm EeV}\,.
\end{equation}
Adopting $B\sim B_{\rm eq}$ is likely too optimistic, but for typical $B$ values in large-scale jets \citep[e.g.,][]{ito21}, and given that $B$ can be enhanced in the ejecta and jet shocks, $0.1\lesssim B/B_{\rm eq}\lesssim 1$ seems plausible. Thus, if efficient acceleration of nuclei occurs in a (mildly) relativistic jet shock, taking $v_{\rm s}\sim 10^{10}$~cm~s$^{-1}$, $R_{\rm ej}\sim r_{\rm j}$ (i.e., an expanded ejecta), and an intermediate $B$ strength $\sim 0.3\,B_{\rm eq}$, $E_{\rm max}$ can reach $\sim 30\,Z_1$~EeV. 

We note that even if $v_{\rm j}\rightarrow c$, limitations of acceleration in relativistic shocks may
not actually apply \citep[see, e.g.,][]{cer23,hua23}. In particular, in the present context,
$\gtrsim$~EeV nuclei may bounce back and forth between the jet upstream and the shocked ejecta
\citep[see][]{bos12}. This mechanism is similar to that studied in other scenarios
\citep[e.g.,][]{byk21,mal23}, and to the espresso mechanism \citep{cap15}, as all of them allow for a
high gain in each shock-crossing and a hard spectrum\footnote{A somewhat similar process expected to
operate in more compact regions is the converter mechanism \citep{der03,ste03}.}. As a result of all of this, we conclude that the jet-ejecta interaction is a potential accelerator of UHECRs and, at the very least, can provide the seeds for UHECR production elsewhere in the jet and its surroundings.

\section{CR injection rates from jet-ejecta interactions}\label{comp}

It is informative to compare the rate of CRs potentially injected directly by SNe inside the jet with that of diffusive entrainment of GCRs. We recall that these processes can occur simultaneously, and do not preclude each other. For this comparison, we assumed co-spatial, uniform, and isotropic distributions of the GCR density and injection rate in a galaxy of radius $R_{\rm gal}$ and (half) height $z_{\rm gal}$. We considered the following as well: the GCR injection rate proportional to the SN rate $\dot N_{\rm SNR}$; galaxy and jet half volumes $V_{\rm gal}\sim \pi R_{\rm gal}^2z_{\rm gal}$ and $V_{\rm j}\sim \pi r_{\rm j}^2z_{\rm gal}$, respectively; and galaxy and jet boundary surfaces for those volumes $S_{\rm gal}\sim \pi R_{\rm gal}^2+2\pi R_{\rm gal}z_{\rm gal}$ (excluding the galaxy mid plane) and $S_{\rm j}\sim 2\pi r_{\rm j}z_{\rm gal}$ (excluding the jet extremes), respectively. We also assumed that GCRs cross the jet-galaxy boundary at the same rate as they cross any other boundary, but we note that the jet-medium shear layer could affect GCR penetration. Finally, taking $z_{\rm gal}=\xi R_{\rm gal}$ and $r_{\rm j}=\theta z_{\rm gal}$ ($\xi\sim 0.1-1$ and $\theta\sim 0.01-0.1$~rad may be reasonable values), one can write the following: $V_{\rm gal}\sim \pi\xi^{-2}z_{\rm gal}^3$;
$V_{\rm j}\sim \pi\theta^2z_{\rm gal}^3$; $S_{\rm gal}\sim \pi\xi^{-2}z_{\rm gal}^2+2\pi \xi^{-1}z_{\rm gal}^2$; and
$S_{\rm j}\sim 2\pi\theta z_{\rm gal}^2$. These assumptions and characterizations are affected by large uncertainties since the individual galaxy and jet properties are very diverse, and jet GCR entrainment and particle acceleration in jet-ejecta interactions are still poorly understood.

The energy rate with which the jet entrains GCRs is
\begin{equation}
\dot E_{\rm GCR}\sim E_{\rm SNR,CR}\dot N_{\rm SNR}(S_{\rm j}/S_{\rm gal})\propto 2\theta/(\xi^{-2}+2\xi^{-1})\lesssim \theta\xi\,,
\label{extr}
\end{equation} 
whereas the rate of CR energy injected by SNe within the jet is
\begin{equation}
\dot E_{\rm s,CR}\sim E_{\rm s,CR}\dot N_{\rm SNR}(V_{\rm j}/V_{\rm gal})\propto (\xi\theta)^2\,,
\label{intr}
\end{equation}
where $E_{\rm SNR,CR}$ and $E_{\rm s,CR}$ are the CR energy released by regular SNRs and jet-ejecta interactions (adding all the stages), respectively. Equation~(\ref{extr}) is strictly valid only if GCRs come from SNRs, but even if not SNRs may still be a good proxy to determine $\dot E_{\rm GCR}$. Equations~(\ref{extr}--\ref{intr}) show that for $E_{\rm s,CR}/E_{\rm SNR,CR}>1/\theta\xi$, SNe in
AGN jets can dominate the injection of UHECR seeds over jet-entrained GCRs. As an example, for $\xi\sim 1$
(an elliptical host) and $\theta\sim 0.1$~rad, even ejecta shocks alone (the total $E_{\rm s}$ is in fact far larger) may surpass the entrained GCR
contribution to the UHECR seeds. 

Taking  $t_{\rm dyn}\sim 10^4$~yr and a rate of SNe inside the jet of $t_{\rm SN,j}^{-1}\sim
10^{-3}\dot N_{\rm SN,1/century}\xi_{-1}^2\theta_{-1}^2$~century$^{-1}$, one obtains a percentage of
$\sim 10$\% of AGN jets hosting some jet-ejecta interaction, with the duty cycle being $D\sim 0.1\,t_{\rm
dyn,11.5}t_{\rm SN,j,12.5}^{-1}$. This allows us to derive the maximum luminosity that can go to CRs in these events, that is, $\dot E_{\rm s,CR}\lesssim 0.1\,D_{-1}L_{\rm j}$. \cite{cap15} estimates that AGNe should produce $\gtrsim 10^{-3}$ of
their bolometric luminosity in the form of CRs to fulfill the energy requirement to be UHECR sources,
and SN-jet interactions appear to be able to fulfill that condition. Predictions for electromagnetic
evidence are beyond the scope of this work, but these events may be hard to disentangle from
other kiloparsec-scale jet dissipative processes; due to moderate $L_{\rm j}$ values and slow radiative
losses, they should be faint neutrino sources.

\section{Conclusions}\label{conc}

Supernovae exploding inside AGN jets can naturally provide both light and heavy seeds to the
processes accelerating UHECRs, which reach Earth with a rather heavy composition above $\sim
10$~EeV. The energy involved in individual jet-ejecta interactions, $E_{\rm s}$, which is an upper-limit to the
CR energy produced by them, can be very high. Considering $M_{\rm ej}=2$~M$_\odot$, $L_{\rm
j,44}=1$, and $E_{\rm SN,51}=1 $ as reference, one gets the following: after an initial $t_{\rm max}\sim 1$~kyr, one expects
$E_{\rm s}\sim 10^{52}$~erg in the ejecta shock, and $E_{\rm s}\sim 10^{53}$~erg in the jet shock;
and after $t_{\rm dyn}\sim 10$~kyr, $E_{\rm s}\sim 10^{55}$~erg, which is associated with the jet pushing on
the ejecta during $t_{\rm dyn}$. Assuming a modest duty cycle of $D\sim 0.01$ and $E_{\rm s,CR}\sim 0.1\,E_{\rm s}$, the total time-averaged, CR luminosity per source is $\sim 10^{41}\,L_{\rm j,44}$~erg~s$^{-1}$. The properties of these interactions seem adequate for the
acceleration of nuclei with $Z\sim 10$ up to $\sim 1$~EeV initially, and $\sim 30$~EeV in later
stages. The luminosities derived for these events could overcome that of jet-entrained GCRs in
pre-accelerated seeds for the production of UHECRs inside the jets, and they may even directly contribute to the observed UHECRs significantly. We remark that rare and strong mass loss
by evolved stars in the jet could present a similar phenomenology to that of SNe. We finish by noting that
powerful and fast winds in non-jetted AGNe \citep[][and references therein]{lah21}
should also host SNe, with a total energy involved in wind-ejecta interactions that may be 
similar to the jet case.


\begin{acknowledgements}
We thank the anonymous referee for constructive and useful comments that really helped to improve the article. We are grateful to A. M. Taylor and D. Khangulyan for their insightful comments on the manuscript.
V.B-R. acknowledges financial support from the State Agency for Research of the Spanish Ministry of Science and Innovation under grants PID2019-105510GB-C31/AEI/10.13039/501100011033/ 
and PID2022-136828NB-C41/AEI/10.13039/501100011033/, and by "ERDF A way of
making Europe" (EU), and through the ''Unit of Excellence Mar\'ia de Maeztu 2020-2023'' award to the Institute of Cosmos Sciences (CEX2019-000918-M). V.B-R. is Correspondent Researcher of CONICET, Argentina, at the IAR.
\end{acknowledgements}

   \bibliographystyle{aa} 
   \bibliography{biblio} 

\begin{thebibliography}{67}
\expandafter\ifx\csname natexlab\endcsname\relax\def\natexlab#1{#1}\fi

\bibitem[{{Aab} {et~al.}(2014){Aab}, {Abreu}, {Aglietta}, {Ahn}, {Al Samarai},
  {Albuquerque}, {Allekotte}, {Allen}, {Allison}, {Almela}, {Alvarez Castillo},
  {Alvarez-Mu{\~n}iz}, {Alves Batista}, {Ambrosio}, {Aminaei}, {Anchordoqui},
  {Andringa}, {Aramo}, {Aranda}, {Arqueros}, {Asorey}, {Assis}, {Aublin},
  {Ave}, {Avenier}, {Avila}, {Awal}, {Badescu}, {Barber}, {B{\"a}uml}, {Baus},
  {Beatty}, {Becker}, {Bellido}, {Berat}, {Bertania}, {Bertou}, {Biermann},
  {Billoir}, {Blaess}, {Blanco}, {Bleve}, {Bl{\"u}mer},
  {Boh{\'a}{\v{c}}ov{\'a}}, {Boncioli}, {Bonifazi}, {Bonino}, {Borodai},
  {Brack}, {Brancus}, {Bridgeman}, {Brogueira}, {Brown}, {Buchholz}, {Bueno},
  {Buitink}, {Buscemi}, {Caballero-Mora}, {Caccianiga}, {Caccianiga},
  {Candusso}, {Caramete}, {Caruso}, {Castellina}, {Cataldi}, {Cazon}, {Cester},
  {Chavez}, {Chiavassa}, {Chinellato}, {Chudoba}, {Cilmo}, {Clay}, {Cocciolo},
  {Colalillo}, {Coleman}, {Collica}, {Coluccia}, {Concei{\c{c}}{\~a}o},
  {Contreras}, {Cooper}, {Cordier}, {Coutu}, {Covault}, {Cronin}, {Curutiu},
  {Dallier}, {Daniel}, {Dasso}, {Daumiller}, {Dawson}, {de Almeida}, {De
  Domenico}, {de Jong}, {de Mello Neto}, {De Mitri}, {de Oliveira}, {de Souza},
  {del Peral}, {Deligny}, {Dembinski}, {Dhital}, {Di Giulio}, {Di Matteo},
  {Diaz}, {D{\'\i}az Castro}, {Diogo}, {Dobrigkeit}, {Docters}, {D'Olivo},
  {Dorofeev}, {Dorosti Hasankiadeh}, {Dova}, {Ebr}, {Engel}, {Erdmann},
  {Erfani}, {Escobar}, {Espadanal}, {Etchegoyen}, {Facal San Luis}, {Falcke},
  {Fang}, {Farrar}, {Fauth}, {Fazzini}, {Ferguson}, {Fernandes}, {Fick},
  {Figueira}, {Filevich}, {Filip{\v{c}}i{\v{c}}}, {Fox}, {Fratu},
  {Fr{\"o}hlich}, {Fuchs}, {Fuji}, {Gaior}, {Garc{\'\i}a}, {Garcia Roca},
  {Garcia-Gamez}, {Garcia-Pinto}, {Garilli}, {Gascon Bravo}, {Gate}, {Gemmeke},
  {Ghia}, {Giaccari}, {Giammarchi}, {Giller}, {Glaser}, {Glass}, {G{\'o}mez
  Berisso}, {G{\'o}mez Vitale}, {Gon{\c{c}}alves}, {Gonzalez}, {Gonz{\'a}lez},
  {Gookin}, {Gordon}, {Gorgi}, {Gorham}, {Gouffon}, {Grebe}, {Griffith},
  {Grillo}, {Grubb}, {Guarino}, {Guedes}, {Hampel}, {Hansen}, {Harari},
  {Harrison}, {Hartmann}, {Harton}, {Haungs}, {Hebbeker}, {Heck}, {Heimann},
  {Herve}, {Hill}, {Hojvat}, {Hollon}, {Holt}, {Homola}, {H{\"o}randel},
  {Horvath}, {Hrabovsk{\'y}}, {Huber}, {Huege}, {Insolia}, {Isar}, {Jandt},
  {Jansen}, {Jarne}, {Josebachuili}, {K{\"a}{\"a}p{\"a}}, {Kambeitz},
  {Kampert}, {Kasper}, {Katkov}, {K{\'e}gl}, {Keilhauer}, {Keivani}, {Kemp},
  {Kieckhafer}, {Klages}, {Kleifges}, {Kleinfeller}, {Krause}, {Krohm},
  {Kr{\"o}mer}, {Kruppke-Hansen}, {Kuempel}, {Kunka}, {LaHurd}, {Latronico},
  {Lauer}, {Lauscher}, {Lautridou}, {Le Coz}, {Le{\~a}o}, {Lebrun}, {Lebrun},
  {Leigui de Oliveira}, {Letessier-Selvon}, {Lhenry-Yvon}, {Link}, {L{\'o}pez},
  {Lopez Ag{\"u}era}, {Louedec}, {Lozano Bahilo}, {Lu}, {Lucero}, {Ludwig},
  {Malacari}, {Maldera}, {Mallamaci}, {Maller}, {Mandat}, {Mantsch},
  {Mariazzi}, {Marin}, {Mari{\c{s}}}, {Marsella}, {Martello}, {Martin},
  {Martinez}, {Mart{\'\i}nez Bravo}, {Martraire}, {Mas{\'\i}as Meza}, {Mathes},
  {Mathys}, {Matthews}, {Matthews}, {Matthiae}, {Maurel}, {Maurizio},
  {Mayotte}, {Mazur}, {Medina}, {Medina-Tanco}, {Meissner}, {Melissas}, {Melo},
  {Menshikov}, {Messina}, {Meyhandan}, {Mi{\'c}anovi{\'c}}, {Micheletti},
  {Middendorf}, {Minaya}, {Miramonti}, {Mitrica}, {Molina-Bueno}, {Mollerach},
  {Monasor}, {Monnier Ragaigne}, {Montanet}, {Morello}, {Mostaf{\'a}}, {Moura},
  {Muller}, {M{\"u}ller}, {M{\"u}ller}, {M{\"u}nchmeyer}, {Mussa}, {Navarra},
  {Navas}, {Necesal}, {Nellen}, {Nelles}, {Neuser}, {Nguyen}, {Niechciol},
  {Niemietz}, {Niggemann}, {Nitz}, {Nosek}, {Novotny}, {No{\v{z}}ka}, {Ochilo},
  {Olinto}, {Oliveira}, {Pacheco}, {Pakk Selmi-Dei}, {Palatka}, {Pallotta},
  {Palmieri}, {Papenbreer}, {Parente}, {Parra}, {Paul}, {Pech}, {P{\k{e}}kala},
  {Pelayo}, {Pepe}, {Perrone}, {Petermann}, {Peters}, {Petrera}, {Petrov},
  {Phuntsok}, {Piegaia}, {Pierog}, {Pieroni}, {Pimenta}, {Pirronello},
  {Platino}, {Plum}, {Porcelli}, {Porowski}, {Prado}, {Privitera}, {Prouza},
  {Purrello}, {Quel}, {Querchfeld}, {Quinn}, {Rautenberg}, {Ravel},
  {Ravignani}, {Revenu}, {Ridky}, {Riggi}, {Risse}, {Ristori}, {Rizi},
  {Rodrigues de Carvalho}, {Rodriguez Cabo}, {Rodriguez Fernandez}, {Rodriguez
  Rojo}, {Rodr{\'\i}guez-Fr{\'\i}as}, {Rogozin}, {Ros}, {Rosado}, {Rossler},
  {Roth}, {Roulet}, {Rovero}, {Saffi}, {Saftoiu}, {Salamida}, {Salazar},
  {Saleh}, {Salesa Greus}, {Salina}, {S{\'a}nchez}, {Sanchez-Lucas}, {Santo},
  {Santos}, {Santos}, {Sarazin}, {Sarkar}, {Sarmento}, {Sato}, {Scharf},
  {Scherini}, {Schieler}, {Schiffer}, {Schmidt}, {Scholten}, {Schoorlemmer},
  {Schov{\'a}nek}, {Schulz}, {Schulz}, {Schumacher}, {Sciutto}, {Segreto},
  {Settimo}, {Shadkam}, {Shellard}, {Sidelnik}, {Sigl}, {Sima},
  {{\'S}mia{\l}kowski}, {{\v{S}}m{\'\i}da}, {Snow}, {Sommers}, {Sorokin},
  {Squartini}, {Srivastava}, {Stani{\v{c}}}, {Stapleton}, {Stasielak},
  {Stephan}, {Stutz}, {Suarez}, {Suomij{\"a}rvi}, {Supanitsky}, {Sutherland},
  {Swain}, {Szadkowski}, {Szuba}, {Taborda}, {Tapia}, {Tartare}, {Tepe},
  {Theodoro}, {Timmermans}, {Todero Peixoto}, {Toma}, {Tomankova}, {Tom{\'e}},
  {Tonachini}, {Torralba Elipe}, {Torres Machado}, {Travnicek}, {Trovato},
  {Tueros}, {Ulrich}, {Unger}, {Urban}, {Vald{\'e}s Galicia}, {Vali{\~n}o},
  {Valore}, {van Aar}, {van Bodegom}, {van den Berg}, {van Velzen}, {van
  Vliet}, {Varela}, {Vargas C{\'a}rdenas}, {Varner}, {V{\'a}zquez},
  {V{\'a}zquez}, {Veberi{\v{c}}}, {Verzi}, {Vicha}, {Videla}, {Villase{\~n}or},
  {Vlcek}, {Vorobiov}, {Wahlberg}, {Wainberg}, {Walz}, {Watson}, {Weber},
  {Weidenhaupt}, {Weindl}, {Werner}, {Widom}, {Wiencke}, {Wilczy{\'n}ska},
  {Wilczy{\'n}ski}, {Will}, {Williams}, {Winchen}, {Wittkowski}, {Wundheiler},
  {Wykes}, {Yamamoto}, {Yapici}, {Yuan}, {Yushkov}, {Zamorano}, {Zas},
  {Zavrtanik}, {Zavrtanik}, {Zaw}, {Zepeda}, {Zhou}, {Zhu}, {Zimbres Silva},
  {Ziolkowski}, {Zuccarello}, \& {Pierre Auger Collaboration}}]{aab14}
{Aab}, A., {Abreu}, P., {Aglietta}, M., {et~al.} 2014, \prd, 90, 122006

\bibitem[{{Abdul Halim} {et~al.}(2023){Abdul Halim}, {Abreu}, {Aglietta},
  {Allekotte}, {Almeida Cheminant}, {Almela}, {Alvarez-Mu{\~n}iz}, {Ammerman
  Yebra}, {Anastasi}, {Anchordoqui}, {Andrada}, {Andringa}, {Aramo},
  {Ara{\'u}jo Ferreira}, {Arnone}, {Arteaga Vel{\'a}zquez}, {Asorey}, {Assis},
  {Avila}, {Avocone}, {Badescu}, {Bakalova}, {Balaceanu}, {Barbato}, {Bellido},
  {Berat}, {Bertaina}, {Bhatta}, {Biermann}, {Binet}, {Bismark}, {Bister},
  {Biteau}, {Blazek}, {Bleve}, {Bl{\"u}mer}, {Boh{\'a}{\v{c}}ov{\'a}},
  {Boncioli}, {Bonifazi}, {Bonneau Arbeletche}, {Borodai}, {Brack}, {Bretz},
  {Brichetto Orchera}, {Briechle}, {Buchholz}, {Bueno}, {Buitink}, {Buscemi},
  {B{\"u}sken}, {Bwembya}, {Caballero-Mora}, {Caccianiga}, {Caracas}, {Caruso},
  {Castellina}, {Catalani}, {Cataldi}, {Cazon}, {Cerda}, {Chinellato},
  {Chudoba}, {Chytka}, {Clay}, {Cobos Cerutti}, {Colalillo}, {Coleman},
  {Coluccia}, {Concei{\c{c}}{\~a}o}, {Condorelli}, {Consolati}, {Conte},
  {Contreras}, {Convenga}, {Correia dos Santos}, {Covault}, {Cristinziani},
  {Cruz Sanchez}, {Dasso}, {Daumiller}, {Dawson}, {de Almeida}, {de Jes{\'u}s},
  {de Jong}, {de Mello Neto}, {De Mitri}, {de Oliveira}, {de Oliveira Franco},
  {de Palma}, {de Souza}, {De Vito}, {Del Popolo}, {Deligny}, {Deval}, {di
  Matteo}, {Dobre}, {Dobrigkeit}, {D'Olivo}, {Domingues Mendes}, {dos Anjos},
  {Ebr}, {Eman}, {Engel}, {Epicoco}, {Erdmann}, {Etchegoyen}, {Falcke},
  {Farmer}, {Farrar}, {Fauth}, {Fazzini}, {Feldbusch}, {Fenu}, {Fick},
  {Figueira}, {Filip{\v{c}}i{\v{c}}}, {Fitoussi}, {Flaggs}, {Fodran}, {Fujii},
  {Fuster}, {Galea}, {Galelli}, {Garc{\'\i}a}, {Gemmeke}, {Gesualdi},
  {Gherghel-Lascu}, {Ghia}, {Giaccari}, {Giammarchi}, {Glombitza}, {Gobbi},
  {Gollan}, {Golup}, {G{\'o}mez Berisso}, {G{\'o}mez Vitale}, {Gongora},
  {Gonz{\'a}lez}, {Gonz{\'a}lez}, {Goos}, {G{\'o}ra}, {Gorgi}, {Gottowik},
  {Grubb}, {Guarino}, {Guedes}, {Guido}, {Hahn}, {Hamal}, {Hampel}, {Hansen},
  {Harari}, {Harvey}, {Haungs}, {Hebbeker}, {Heck}, {Hojvat}, {H{\"o}randel},
  {Horvath}, {Hrabovsk{\'y}}, {Huege}, {Insolia}, {Isar}, {Janecek}, {Johnsen},
  {Jurysek}, {K{\"a}{\"a}p{\"a}}, {Kampert}, {Keilhauer}, {Khakurdikar},
  {Kizakke Covilakam}, {Klages}, {Kleifges}, {Kleinfeller}, {Knapp}, {Kunka},
  {Lago}, {Langner}, {Leigui de Oliveira}, {Lenok}, {Letessier-Selvon},
  {Lhenry-Yvon}, {Lo Presti}, {Lopes}, {L{\'o}pez}, {Lu}, {Luce}, {Lundquist},
  {Machado Payeras}, {Majercakova}, {Mandat}, {Manning}, {Manshanden},
  {Mantsch}, {Marafico}, {Mariani}, {Mariazzi}, {Mari{\c{s}}}, {Marsella},
  {Martello}, {Martinelli}, {Mart{\'\i}nez Bravo}, {Martins}, {Mastrodicasa},
  {Mathes}, {Matthews}, {Matthiae}, {Mayotte}, {Mayotte}, {Mazur},
  {Medina-Tanco}, {Meinert}, {Melo}, {Menshikov}, {Michal}, {Micheletti},
  {Miramonti}, {Mollerach}, {Montanet}, {Morejon}, {Morello}, {M{\"u}ller},
  {Mulrey}, {Mussa}, {Muzio}, {Namasaka}, {Nasr-Esfahani}, {Nellen}, {Nicora},
  {Niculescu-Oglinzanu}, {Niechciol}, {Nitz}, {Norwood}, {Nosek}, {Novotny},
  {No{\v{z}}ka}, {Nucita}, {N{\'u}{\~n}ez}, {Oliveira}, {Palatka}, {Pallotta},
  {Parente}, {Parra}, {Pawlowsky}, {Pech}, {P{\c{e}}kala}, {Pelayo}, {Pereira
  Martins}, {Perez Armand}, {P{\'e}rez Bertolli}, {Perrone}, {Petrera},
  {Petrucci}, {Pierog}, {Pimenta}, {Platino}, {Pont}, {Pothast}, {Pourmohammad
  Shavar}, {Privitera}, {Prouza}, {Puyleart}, {Querchfeld}, {Rautenberg},
  {Ravignani}, {Reininghaus}, {Ridky}, {Riehn}, {Risse}, {Rizi}, {Rodrigues de
  Carvalho}, {Rodriguez Rojo}, {Roncoroni}, {Rossoni}, {Roth}, {Roulet},
  {Rovero}, {Ruehl}, {Saftoiu}, {Saharan}, {Salamida}, {Salazar}, {Salina},
  {Sanabria Gomez}, {S{\'a}nchez}, {Santos}, {Santos}, {Sarazin}, {Sarmento},
  {Sato}, {Savina}, {Sch{\"a}fer}, {Scherini}, {Schieler}, {Schimassek},
  {Schimp}, {Schl{\"u}ter}, {Schmidt}, {Scholten}, {Schoorlemmer},
  {Schov{\'a}nek}, {Schr{\"o}der}, {Schulte}, {Schulz}, {Sciutto},
  {Scornavacche}, {Segreto}, {Sehgal}, {Shivashankara}, {Sigl}, {Silli},
  {Sima}, {Smau}, {{\v{S}}m{\'\i}da}, {Sommers}, {Soriano}, {Squartini},
  {Stadelmaier}, {Stanca}, {Stani{\v{c}}}, {Stasielak}, {Stassi}, {Straub},
  {Streich}, {Su{\'a}rez-Dur{\'a}n}, {Suomij{\"a}rvi}, {Supanitsky},
  {Szadkowski}, {Tapia}, {Taricco}, {Timmermans}, {Tkachenko}, {Tobiska},
  {Todero Peixoto}, {Tom{\'e}}, {Torr{\`e}s}, {Travaini}, {Travnicek},
  {Trimarelli}, {Tueros}, {Ulrich}, {Unger}, {Vaclavek}, {Vacula}, {Vald{\'e}s
  Galicia}, {Valore}, {Varela}, {V{\'a}squez-Ram{\'\i}rez}, {Veberi{\v{c}}},
  {Ventura}, {Vergara Quispe}, {Verzi}, {Vicha}, {Vink}, {Vorobiov},
  {Watanabe}, {Watson}, {Weindl}, {Wiencke}, {Wilczy{\'n}ski}, {Wittkowski},
  {Wundheiler}, {Yushkov}, {Zapparrata}, {Zas}, {Zavrtanik}, {Zavrtanik}, \&
  {Pierre Auger Collaboration}}]{abd23}
{Abdul Halim}, A., {Abreu}, P., {Aglietta}, M., {et~al.} 2023, \jcap, 2023, 024

\bibitem[{{Aharonian}(2002)}]{aha02b}
{Aharonian}, F.~A. 2002, \mnras, 332, 215

\bibitem[{{Aharonian} {et~al.}(2002){Aharonian}, {Belyanin}, {Derishev},
  {Kocharovsky}, \& {Kocharovsky}}]{aha02}
{Aharonian}, F.~A., {Belyanin}, A.~A., {Derishev}, E.~V., {Kocharovsky}, V.~V.,
  \& {Kocharovsky}, V.~V. 2002, \prd, 66, 023005

\bibitem[{{Anchordoqui}(2018)}]{anc18}
{Anchordoqui}, L.~A. 2018, \prd, 97, 063010

\bibitem[{{Araudo} {et~al.}(2013){Araudo}, {Bosch-Ramon}, \& {Romero}}]{ara13}
{Araudo}, A.~T., {Bosch-Ramon}, V., \& {Romero}, G.~E. 2013, \mnras, 436, 3626

\bibitem[{{Barkov} {et~al.}(2012){Barkov}, {Aharonian}, {Bogovalov}, {Kelner},
  \& {Khangulyan}}]{bar12}
{Barkov}, M.~V., {Aharonian}, F.~A., {Bogovalov}, S.~V., {Kelner}, S.~R., \&
  {Khangulyan}, D. 2012, \apj, 749, 119

\bibitem[{{Barkov} {et~al.}(2010){Barkov}, {Aharonian}, \&
  {Bosch-Ramon}}]{bar10}
{Barkov}, M.~V., {Aharonian}, F.~A., \& {Bosch-Ramon}, V. 2010, \apj, 724, 1517

\bibitem[{{Bednarek}(1999)}]{bed99}
{Bednarek}, W. {\l}.~O. 1999, in Plasma Turbulence and Energetic Particles in
  Astrophysics, ed. M.~{Ostrowski} \& R.~{Schlickeiser}, 360--365

\bibitem[{{Blandford} \& {Koenigl}(1979)}]{bla79}
{Blandford}, R.~D. \& {Koenigl}, A. 1979, \aplett, 20, 15

\bibitem[{{Bosch-Ramon}(2012)}]{bos12}
{Bosch-Ramon}, V. 2012, \aap, 542, A125

\bibitem[{{Burbidge}(1962)}]{bur62}
{Burbidge}, G. 1962, Progress of Theoretical Physics, 27, 999

\bibitem[{{Bykov} {et~al.}(2021){Bykov}, {Petrov}, {Kalyashova}, \&
  {Troitsky}}]{byk21}
{Bykov}, A.~M., {Petrov}, A.~E., {Kalyashova}, M.~E., \& {Troitsky}, S.~V.
  2021, \apjl, 921, L10

\bibitem[{{Caprioli}(2015)}]{cap15}
{Caprioli}, D. 2015, \apjl, 811, L38

\bibitem[{{Cerutti} \& {Giacinti}(2023)}]{cer23}
{Cerutti}, B. \& {Giacinti}, G. 2023, \aap, 676, A23

\bibitem[{{Cronin} {et~al.}(2021){Cronin}, {Utomo}, {Leroy}, {Behrens},
  {Chastenet}, {Holland-Ashford}, {Koch}, {Lopez}, {Sandstrom}, \&
  {Williams}}]{cro21}
{Cronin}, S.~A., {Utomo}, D., {Leroy}, A.~K., {et~al.} 2021, \apj, 923, 86

\bibitem[{{Dayal} \& {Ferrara}(2018)}]{day18}
{Dayal}, P. \& {Ferrara}, A. 2018, \physrep, 780, 1

\bibitem[{{Derishev} {et~al.}(2003){Derishev}, {Aharonian}, {Kocharovsky}, \&
  {Kocharovsky}}]{der03}
{Derishev}, E.~V., {Aharonian}, F.~A., {Kocharovsky}, V.~V., \& {Kocharovsky},
  V.~V. 2003, \prd, 68, 043003

\bibitem[{{Dermer}(2007)}]{der07}
{Dermer}, C.~D. 2007, arXiv e-prints, arXiv:0711.2804

\bibitem[{{di Matteo} {et~al.}(2023){di Matteo}, {Anchordoqui}, {Bister}, {de
  Almeida}, {Deligny}, {Deval}, {Farrar}, {Giaccari}, {Golup}, {Higuchi},
  {Kim}, {Kuznetsov}, {Mari{\c{s}}}, {Rubtsov}, \& {Tinyakov}}]{dim23}
{di Matteo}, A., {Anchordoqui}, L., {Bister}, T., {et~al.} 2023, in European
  Physical Journal Web of Conferences, Vol. 283, European Physical Journal Web
  of Conferences, 03002

\bibitem[{{Fedorenko} \& {Courvoisier}(1996)}]{fed96}
{Fedorenko}, V.~N. \& {Courvoisier}, T.~J.~L. 1996, \aap, 307, 347

\bibitem[{{Gebhardt} \& {Thomas}(2009)}]{geb09}
{Gebhardt}, K. \& {Thomas}, J. 2009, \apj, 700, 1690

\bibitem[{{Greisen}(1966)}]{gre66}
{Greisen}, K. 1966, \prl, 16, 748

\bibitem[{{Hakobyan} {et~al.}(2017){Hakobyan}, {Barkhudaryan}, {Karapetyan},
  {Mamon}, {Kunth}, {Adibekyan}, {Aramyan}, {Petrosian}, \& {Turatto}}]{hak17}
{Hakobyan}, A.~A., {Barkhudaryan}, L.~V., {Karapetyan}, A.~G., {et~al.} 2017,
  \mnras, 471, 1390

\bibitem[{{Hillas}(1984)}]{hil84}
{Hillas}, A.~M. 1984, \araa, 22, 425

\bibitem[{{Huang} {et~al.}(2023){Huang}, {Reville}, {Kirk}, \&
  {Giacinti}}]{hua23}
{Huang}, Z.-Q., {Reville}, B., {Kirk}, J.~G., \& {Giacinti}, G. 2023, \mnras,
  522, 4955

\bibitem[{{Hubbard} \& {Blackman}(2006)}]{hub06}
{Hubbard}, A. \& {Blackman}, E.~G. 2006, \mnras, 371, 1717

\bibitem[{{Ito} {et~al.}(2021){Ito}, {Inoue}, \& {Kataoka}}]{ito21}
{Ito}, S., {Inoue}, Y., \& {Kataoka}, J. 2021, \apj, 916, 95

\bibitem[{{Kelner} \& {Aharonian}(2008)}]{kel08}
{Kelner}, S.~R. \& {Aharonian}, F.~A. 2008, \prd, 78, 034013

\bibitem[{{Khangulyan} {et~al.}(2013){Khangulyan}, {Barkov}, {Bosch-Ramon},
  {Aharonian}, \& {Dorodnitsyn}}]{kha13}
{Khangulyan}, D.~V., {Barkov}, M.~V., {Bosch-Ramon}, V., {Aharonian}, F.~A., \&
  {Dorodnitsyn}, A.~V. 2013, \apj, 774, 113

\bibitem[{{Kimura} {et~al.}(2018){Kimura}, {Murase}, \& {Zhang}}]{kim18}
{Kimura}, S.~S., {Murase}, K., \& {Zhang}, B.~T. 2018, \prd, 97, 023026

\bibitem[{{Krti{\v{c}}ka}(2014)}]{krt14}
{Krti{\v{c}}ka}, J. 2014, \aap, 564, A70

\bibitem[{{Laha} {et~al.}(2021){Laha}, {Reynolds}, {Reeves}, {Kriss},
  {Guainazzi}, {Smith}, {Veilleux}, \& {Proga}}]{lah21}
{Laha}, S., {Reynolds}, C.~S., {Reeves}, J., {et~al.} 2021, Nature Astronomy,
  5, 13

\bibitem[{{Leahy}(2017)}]{lea17}
{Leahy}, D.~A. 2017, \apj, 837, 36

\bibitem[{{Leahy} \& {Filipovi{\'c}}(2022)}]{lea22}
{Leahy}, D.~A. \& {Filipovi{\'c}}, M.~D. 2022, \apj, 931, 20

\bibitem[{{Lemoine}(2009)}]{lem09}
{Lemoine}, M. 2009, Nuclear Physics B Proceedings Supplements, 190, 174

\bibitem[{{Liu} {et~al.}(2023){Liu}, {R{\"o}pke}, \& {Han}}]{liu23}
{Liu}, Z.-W., {R{\"o}pke}, F.~K., \& {Han}, Z. 2023, Research in Astronomy and
  Astrophysics, 23, 082001

\bibitem[{{Malkov} \& {Lemoine}(2023)}]{mal23}
{Malkov}, M. \& {Lemoine}, M. 2023, \pre, 107, 025201

\bibitem[{{Matthews} {et~al.}(2019){Matthews}, {Bell}, {Blundell}, \&
  {Araudo}}]{mat19}
{Matthews}, J.~H., {Bell}, A.~R., {Blundell}, K.~M., \& {Araudo}, A.~T. 2019,
  \mnras, 482, 4303

\bibitem[{{Matthews} \& {Taylor}(2023)}]{mat23}
{Matthews}, J.~H. \& {Taylor}, A.~M. 2023, arXiv e-prints, arXiv:2301.02682

\bibitem[{{Mbarek} \& {Caprioli}(2019)}]{mba19}
{Mbarek}, R. \& {Caprioli}, D. 2019, \apj, 886, 8

\bibitem[{{Mingo} {et~al.}(2019){Mingo}, {Croston}, {Hardcastle}, {Best},
  {Duncan}, {Morganti}, {Rottgering}, {Sabater}, {Shimwell}, {Williams},
  {Brienza}, {Gurkan}, {Mahatma}, {Morabito}, {Prandoni}, {Bondi}, {Ineson}, \&
  {Mooney}}]{min19}
{Mingo}, B., {Croston}, J.~H., {Hardcastle}, M.~J., {et~al.} 2019, \mnras, 488,
  2701

\bibitem[{{Mullin} \& {Hardcastle}(2009)}]{mul09}
{Mullin}, L.~M. \& {Hardcastle}, M.~J. 2009, \mnras, 398, 1989

\bibitem[{{Ostrowski}(1998)}]{ost98}
{Ostrowski}, M. 1998, \aap, 335, 134

\bibitem[{{Perucho}(2019)}]{per19}
{Perucho}, M. 2019, Galaxies, 7, 70

\bibitem[{{Perucho} {et~al.}(2017){Perucho}, {Bosch-Ramon}, \&
  {Barkov}}]{per17}
{Perucho}, M., {Bosch-Ramon}, V., \& {Barkov}, M.~V. 2017, \aap, 606, A40

\bibitem[{{Perucho} {et~al.}(2014){Perucho}, {Mart{\'\i}}, {Laing}, \&
  {Hardee}}]{per14}
{Perucho}, M., {Mart{\'\i}}, J.~M., {Laing}, R.~A., \& {Hardee}, P.~E. 2014,
  \mnras, 441, 1488

\bibitem[{{Petruk} {et~al.}(2021){Petruk}, {Kuzyo}, {Orlando}, {Pohl}, \&
  {Brose}}]{pet21}
{Petruk}, O., {Kuzyo}, T., {Orlando}, S., {Pohl}, M., \& {Brose}, R. 2021,
  \mnras, 505, 755

\bibitem[{{Rachen}(1996)}]{rac96}
{Rachen}, J.~P. 1996, PhD thesis, Max-Planck-Institute for Radioastronomy, Bonn

\bibitem[{{Rachen} \& {Biermann}(1993)}]{rac93}
{Rachen}, J.~P. \& {Biermann}, P.~L. 1993, \aap, 272, 161

\bibitem[{{Reddy} {et~al.}(2023){Reddy}, {Georganopoulos}, {Meyer}, {Keenan},
  \& {Kollmann}}]{red23}
{Reddy}, K., {Georganopoulos}, M., {Meyer}, E.~T., {Keenan}, M., \& {Kollmann},
  K.~E. 2023, \apjs, 265, 8

\bibitem[{{Rieger}(2022)}]{rie22}
{Rieger}, F.~M. 2022, Universe, 8, 607

\bibitem[{{Romero} {et~al.}(1996){Romero}, {Combi}, {Perez Bergliaffa}, \&
  {Anchordoqui}}]{rom96}
{Romero}, G.~E., {Combi}, J.~A., {Perez Bergliaffa}, S.~E., \& {Anchordoqui},
  L.~A. 1996, Astroparticle Physics, 5, 279

\bibitem[{{Romero} {et~al.}(2018){Romero}, {M{\"u}ller}, \& {Roth}}]{rom18}
{Romero}, G.~E., {M{\"u}ller}, A.~L., \& {Roth}, M. 2018, \aap, 616, A57

\bibitem[{{Seo} {et~al.}(2023){Seo}, {Ryu}, \& {Kang}}]{seo22}
{Seo}, J., {Ryu}, D., \& {Kang}, H. 2023, \apj, 944, 199

\bibitem[{{Stern}(2003)}]{ste03}
{Stern}, B.~E. 2003, \mnras, 345, 590

\bibitem[{{Taylor} {et~al.}(2015){Taylor}, {Ahlers}, \& {Hooper}}]{tay15}
{Taylor}, A.~M., {Ahlers}, M., \& {Hooper}, D. 2015, \prd, 92, 063011

\bibitem[{{Taylor} {et~al.}(2023){Taylor}, {Matthews}, \& {Bell}}]{tay23}
{Taylor}, A.~M., {Matthews}, J.~H., \& {Bell}, A.~R. 2023, \mnras
  [\eprint[arXiv]{2302.06489}]

\bibitem[{{The Pierre Auger Collaboration} {et~al.}(2019){The Pierre Auger
  Collaboration}, {Aab}, {Abreu}, {Aglietta}, {Albuquerque}, {Albury},
  {Allekotte}, {Almela}, {Alvarez Castillo}, {Alvarez-Mu{\~n}iz}, {Anastasi},
  {Anchordoqui}, {Andrada}, {Andringa}, {Aramo}, {Asorey}, {Assis}, {Avila},
  {Badescu}, {Bakalova}, {Balaceanu}, {Barbato}, {Barreira Luz}, {Baur},
  {Becker}, {Bellido}, {Berat}, {Bertaina}, {Bertou}, {Biermann}, {Bister},
  {Biteau}, {Blanco}, {Blazek}, {Bleve}, {Boh{\'a}{\v{c}}ov{\'a}}, {Boncioli},
  {Bonifazi}, {Borodai}, {Botti}, {Brack}, {Bretz}, {Briechle}, {Buchholz},
  {Bueno}, {Buitink}, {Buscemi}, {Caballero-Mora}, {Caccianiga}, {Calcagni},
  {Cancio}, {Canfora}, {Caracas}, {Carceller}, {Caruso}, {Castellina},
  {Catalani}, {Cataldi}, {Cazon}, {Cerda}, {Chinellato}, {Choi}, {Chudoba},
  {Chytka}, {Clay}, {Cobos Cerutti}, {Colalillo}, {Coleman}, {Coluccia},
  {Concei{\c{c}}{\~a}o}, {Condorelli}, {Consolati}, {Contreras}, {Convenga},
  {Cooper}, {Coutu}, {Covault}, {Daniel}, {Dasso}, {Daumiller}, {Dawson},
  {Day}, {de Almeida}, {de Jong}, {De Mauro}, {de Mello Neto}, {De Mitri}, {de
  Oliveira}, {de Souza}, {Debatin}, {del R{\'\i}o}, {Deligny}, {Dhital}, {Di
  Matteo}, {D{\'\i}az Castro}, {Dobrigkeit}, {D'Olivo}, {Dorosti}, {dos Anjos},
  {Dova}, {Dundovic}, {Ebr}, {Engel}, {Erdmann}, {Escobar}, {Etchegoyen},
  {Falcke}, {Farmer}, {Farrar}, {Fauth}, {Fazzini}, {Feldbusch}, {Fenu},
  {Ferreyro}, {Figueira}, {Filip{\v{c}}i{\v{c}}}, {Freire}, {Fujii}, {Fuster},
  {Garc{\'\i}a}, {Gemmeke}, {Gesualdi}, {Gherghel-Lascu}, {Ghia}, {Giaccari},
  {Giammarchi}, {Giller}, {G{\l}as}, {Glombitza}, {Gobbi}, {Golup}, {G{\'o}mez
  Berisso}, {G{\'o}mez Vitale}, {Gongora}, {Gonz{\'a}lez}, {Goos}, {G{\'o}ra},
  {Gorgi}, {Gottowik}, {Grubb}, {Guarino}, {Guedes}, {Guido}, {Hahn},
  {Halliday}, {Hampel}, {Hansen}, {Harari}, {Harrison}, {Harvey}, {Haungs},
  {Hebbeker}, {Heck}, {Hill}, {Hojvat}, {Holt}, {H{\"o}randel}, {Horvath},
  {Hrabovsk{\'y}}, {Huege}, {Hulsman}, {Insolia}, {Isar}, {Johnsen}, {Jurysek},
  {K{\"a}{\"a}p{\"a}}, {Kampert}, {Keilhauer}, {Kemmerich}, {Kemp}, {Klages},
  {Kleifges}, {Kleinfeller}, {Kukec Mezek}, {Kuotb Awad}, {Lago}, {LaHurd},
  {Lang}, {Legumina}, {Leigui de Oliveira}, {Lenok}, {Letessier-Selvon},
  {Lhenry-Yvon}, {Lippmann}, {Lo Presti}, {Lopes}, {L{\'o}pez}, {L{\'o}pez
  Casado}, {Lorek}, {Luce}, {Lucero}, {Malacari}, {Mancarella}, {Mandat},
  {Manning}, {Manshanden}, {Mantsch}, {Mariazzi}, {Mari{\c{s}}}, {Marsella},
  {Martello}, {Martinez}, {Mart{\'\i}nez Bravo}, {Mastrodicasa}, {Mathes},
  {Mathys}, {Matthews}, {Matthiae}, {Mayotte}, {Mazur}, {Medina-Tanco}, {Melo},
  {Menshikov}, {Merenda}, {Michal}, {Micheletti}, {Miramonti}, {Mockler},
  {Mollerach}, {Montanet}, {Morello}, {Morlino}, {Mostaf{\'a}}, {M{\"u}ller},
  {Muller}, {M{\"u}ller}, {Mussa}, {Namasaka}, {Nellen}, {Niculescu-Oglinzanu},
  {Niechciol}, {Nitz}, {Nosek}, {Novotny}, {No{\v{z}}ka}, {Nucita},
  {N{\'u}{\~n}ez}, {Olinto}, {Palatka}, {Pallotta}, {Panetta}, {Papenbreer},
  {Parente}, {Parra}, {Pech}, {Pedreira}, {P{\k{e}}kala}, {Pelayo},
  {Pe{\~n}a-Rodriguez}, {Pereira}, {Perlin}, {Perrone}, {Peters}, {Petrera},
  {Phuntsok}, {Pierog}, {Pimenta}, {Pirronello}, {Platino}, {Poh}, {Pont},
  {Pothast}, {Prado}, {Privitera}, {Prouza}, {Puyleart}, {Querchfeld},
  {Ramos-Pollan}, {Rautenberg}, {Ravignani}, {Reininghaus}, {Ridky}, {Riehn},
  {Risse}, {Ristori}, {Rizi}, {Rodrigues de Carvalho}, {Rodriguez Rojo},
  {Roncoroni}, {Roth}, {Roulet}, {Rovero}, {Ruehl}, {Saffi}, {Saftoiu},
  {Salamida}, {Salazar}, {Salina}, {Sanabria Gomez}, {S{\'a}nchez}, {Santos},
  {Santos}, {Sarazin}, {Sarmento}, {Sarmiento-Cano}, {Sato}, {Savina},
  {Schauer}, {Scherini}, {Schieler}, {Schimassek}, {Schimp}, {Schl{\"u}ter},
  {Schmidt}, {Scholten}, {Schov{\'a}nek}, {Schr{\"o}der}, {Schr{\"o}der},
  {Schumacher}, {Sciutto}, {Scornavacche}, {Shellard}, {Sigl}, {Silli}, {Sima},
  {{\v{S}}m{\'\i}da}, {Snow}, {Sommers}, {Soriano}, {Souchard}, {Squartini},
  {Stadelmaier}, {Stanca}, {Stani{\v{c}}}, {Stasielak}, {Stassi},
  {Stolpovskiy}, {Streich}, {Su{\'a}rez-Dur{\'a}n}, {Sudholz},
  {Suomij{\"a}rvi}, {Supanitsky}, {{\v{S}}up{\'\i}k}, {Szadkowski}, {Taboada},
  {Taborda}, {Tapia}, {Timmermans}, {Tobiska}, {Todero Peixoto}, {Tom{\'e}},
  {Torralba Elipe}, {Travaini}, {Travnicek}, {Trini}, {Tueros}, {Ulrich},
  {Unger}, {Urban}, {Vald{\'e}s Galicia}, {Vali{\~n}o}, {Valore}, {van den
  Berg}, {van Vliet}, {Varela}, {Vargas C{\'a}rdenas},
  {V{\'a}squez-Ram{\'\i}rez}, {Veberi{\v{c}}}, {Ventura}, {Vergara Quispe},
  {Verzi}, {Vicha}, {Villase{\~n}or}, {Vink}, {Vorobiov}, {Wahlberg}, {Watson},
  {Weber}, {Weindl}, {Wiede{\'n}ski}, {Wiencke}, {Wilczy{\'n}ski}, {Winchen},
  {Wirtz}, {Wittkowski}, {Wundheiler}, {Yang}, {Yushkov}, {Zas}, {Zavrtanik},
  {Zavrtanik}, {Zehrer}, {Zepeda}, {Zimmermann}, {Ziolkowski}, \&
  {Zuccarello}}]{pie19}
{The Pierre Auger Collaboration}, {Aab}, A., {Abreu}, P., {et~al.} 2019, arXiv
  e-prints, arXiv:1909.09073

\bibitem[{{Torres-Alb{\`a}}(2019)}]{tor19}
{Torres-Alb{\`a}}, N. 2019, in High Energy Phenomena in Relativistic
  OutflowsVII, 5

\bibitem[{{Torres-Alb{\`a}} \& {Bosch-Ramon}(2019)}]{tor19b}
{Torres-Alb{\`a}}, N. \& {Bosch-Ramon}, V. 2019, \aap, 623, A91

\bibitem[{{Vieyro} {et~al.}(2019){Vieyro}, {Bosch-Ramon}, \&
  {Torres-Alb{\`a}}}]{vie19}
{Vieyro}, F.~L., {Bosch-Ramon}, V., \& {Torres-Alb{\`a}}, N. 2019, \aap, 622,
  A175

\bibitem[{{Wykes} {et~al.}(2013){Wykes}, {Croston}, {Hardcastle}, {Eilek},
  {Biermann}, {Achterberg}, {Bray}, {Lazarian}, {Haverkorn}, {Protheroe}, \&
  {Bromberg}}]{wyk13}
{Wykes}, S., {Croston}, J.~H., {Hardcastle}, M.~J., {et~al.} 2013, \aap, 558,
  A19

\bibitem[{{Wykes} {et~al.}(2015){Wykes}, {Hardcastle}, {Karakas}, \&
  {Vink}}]{wyk15}
{Wykes}, S., {Hardcastle}, M.~J., {Karakas}, A.~I., \& {Vink}, J.~S. 2015,
  \mnras, 447, 1001

\bibitem[{{Wykes} {et~al.}(2018){Wykes}, {Taylor}, {Bray}, {Hardcastle}, \&
  {Hillas}}]{wyk18}
{Wykes}, S., {Taylor}, A.~M., {Bray}, J.~D., {Hardcastle}, M.~J., \& {Hillas},
  M. 2018, Nuclear and Particle Physics Proceedings, 297-299, 234

\bibitem[{{Zatsepin} \& {Kuz'min}(1966)}]{zat66}
{Zatsepin}, G.~T. \& {Kuz'min}, V.~A. 1966, Soviet Journal of Experimental and
  Theoretical Physics Letters, 4, 78

\bibitem[{{Zirakashvili} {et~al.}(2023){Zirakashvili}, {Ptuskin}, \&
  {Rogovaya}}]{zir23}
{Zirakashvili}, V.~N., {Ptuskin}, V.~S., \& {Rogovaya}, S.~I. 2023, \mnras,
  519, L5

\end{thebibliography}

\end{document}